\documentclass[aps,prl,showpacs,twocolumn,superscriptaddress,
	      floats,floatfix,preprintnumbers]{revtex4}
\usepackage{amssymb,amsmath}
\usepackage{bm,url}
\usepackage{graphics}
\usepackage{subfigure}
\usepackage{epsfig}
\usepackage{ulem}
\usepackage{color}
\usepackage{soul}
\sethlcolor{yellow}
\setulcolor{blue}
\setstcolor{red}
\input{symdef.inc.two}
\definecolor{darkgreen}{rgb}{0,0.625,0}

\begin{document}
\title{Statistical Properties of the Intrinsic Geometry of 
Heavy-particle Trajectories in Two-dimensional, Homogeneous,
Isotropic Turbulence}
\author{Anupam Gupta} 
\email{anupam1509@gmail.com}
\affiliation{ Department of Physics, University of “Tor Vergata”, Via della Ricerca 
Scientifica 1, 00133 Rome, Italy}
\affiliation{Centre for Condensed Matter Theory, Department of Physics, Indian
Institute of Science, Bangalore 560012, India}
\author{Dhrubaditya Mitra} 
\email{dhruba.mitra@gmail.com}
\affiliation{NORDITA, Royal Institute of Technology and Stockholm University,
Roslagstullsbacken 23, SE-10691 Stockholm, Sweden}
\author{Prasad Perlekar}
\email{perlekar@tifrh.res.in}
\affiliation{TIFR Centre for Interdisciplinary Sciences, 21 Brundavan Colony, 
Narsingi, Hyderabad 500075, India}
\author{Rahul Pandit}
\email{rahul@physics.iisc.ernet.in}
\altaffiliation[\\ also at~]{Jawaharlal Nehru Centre For Advanced
Scientific Research, Jakkur, Bangalore, India.}
\affiliation{Centre for Condensed Matter Theory, Department of Physics, Indian
Institute of Science, Bangalore 560012, India} 
\begin{abstract}
We obtain, by extensive direct numerical simulations, trajectories of heavy
inertial particles in two-dimensional, statistically steady, homogeneous, and
isotropic turbulent flows, with friction. We show that the probability
distribution function $\mP(\kappa)$, of the trajectory curvature $\kappa$, is
such that, as $\kappa \to \infty$, $\mP(\kappa) \sim \kappa^{-\hr}$, with  $\hr
= 2.07 \pm 0.09$. The exponent $\hr$ is universal, insofar as it is independent of the Stokes
number $\St$ and the energy-injection wave number $\kinj$. We show that this
exponent lies within error bars of their counterparts for trajectories of
Lagrangian tracers. We demonstrate that the complexity of heavy-particle
trajectories can be characterized by the number $\NI(t,\St)$ of inflection
points (up until time $t$) in the trajectory and $\nI(\St) \equiv
\lim_{t\to\infty} \frac{\NI(t,\St)}{t} \sim\St^{-\Delta}$, where the exponent
$\Delta = 0.33 \pm0.02$ is also universal.  
\end{abstract} 
\preprint{NORDITA-2014-21}

\keywords{turbulence, inertial particle, statistical mechanics} 
\pacs{47.27.-i,05.40.-a } 

\maketitle

The transport of particles by turbulent fluids has attracted considerable
attention since the pioneering work of Taylor~\cite{tay22}.  The study of such
transport has experienced a renaissance because (a) there have been tremendous
advances in measurement techniques and direct numerical simulations
(DNSs)~\cite{tos+bod09} and (b) it has implications not only for fundamental
problems in the physics of turbulence~\cite{bec+bif+bof+cen+mus+tos06} but also
for a variety of geophysical, atmospheric, astrophysical, and industrial
problems~\cite{sha03,gra+wan13,fal+fou+ste02,Arm10,Csa73,eat+fes94,pos+abr02}.
It is natural to use the Lagrangian frame of reference~\cite{fal+gaw+var01}
here; but we must distinguish between (a) Lagrangian or tracer particles, which
are neutrally buoyant and follow the flow velocity at a point, and (b) inertial
particles, whose density $\rho_p$ is different from the density $\rho_f$ of the
advecting fluid. The motion of heavy inertial particles is determined by the
flow drag, which can be parameterized by a time scale $\taus$, whose ratio with
the Kolmogorov dissipation time $\Teta$ is the Stokes number $\St =
\taus/\Teta$; tracer and heavy inertial particles show qualitatively different
behaviors in flows; e.g., the former are uniformly dispersed in a turbulent
flow, whereas the latter cluster~\cite{bec+bif+bof+cen+mus+tos06}, most
prominently when $\St \simeq 1$.  Differences between tracers and inertial
particles have been investigated in several studies~\cite{tos+bod09}, which
have concentrated on three-dimensional (3D) flows and on the clustering or
dispersion of these particles.

We present the first study of the statistical properties of the geometries of
heavy-particle trajectories in two-dimensional (2D), homogeneous, isotropic,
and statistically steady turbulence, which is qualitatively different from its
3D counterpart because, if energy is injected at wave number $\kinj$, two
power-law regimes appear in the energy spectrum
$E(k)$~\cite{kra+mon80,pan+per+ray09,bof+eck12}, for wave numbers $k<\kinj$ and
$k> \kinj$. One regime is associated with an inverse cascade of energy, towards
large length scales, and the other with a forward cascade of enstrophy to small
length scales. It is important to study both forward- and inverse-cascade
regimes, so we use $\kinj = 4$, which gives a large forward-cascade regime in
$E(k)$, and $\kinj=50$, which yields both forward- and inverse-cascade regimes.

For a heavy inertial particle, we calculate the velocity $\vv$, the
acceleration $\aa=d\vv/dt$, with magnitude $a$ and normal and tangential
components $\an$ and $\at$, respectively.  The intrinsic curvature of a
particle trajectory is $\kappa=\an/v^2$. We find two intriguing results that
shed new light on the geometries of particle tracks in 2D turbulence: First,
the probability distribution function (PDF) $\mP(\kappa)$ is such that, as
$\kappa \to \infty$, $\mP(\kappa) \sim \kappa^{-\hr}$; in contrast, as $\kappa
\to 0$, $\mP(\kappa)$ has slope zero; we find that $\hr = 2.07 \pm 0.09$ is
universal, insofar as they are independent of $\St$ and $\kinj$. We present
high-quality data, with \textit{two decades} of clean scaling, to obtain the
values of these exponents, for different values of $\St$ and $\kinj$. We obtain
data of similar quality for Lagrangian-tracer trajectories and thus show that
$\hr$ lies within error bars of its tracer-particle counterpart. Second, along
every heavy-particle track, we calculate the number, $\NI(t,\St)$, of
inflection points (at which $\aa \times \vv$ changes sign) up until time $t$.
We propose that 
\begin{equation}
\nI(\St) \equiv \lim_{t\to\infty} \frac{\NI(t,\St)}{t}
\label{eq:nI}
\end{equation}
is a natural measure of the complexity of the trajectories of these particles;
and we find that $\nI\sim\St^{-\Delta}$, where the exponent $\Delta =
0.33\pm0.02$ is also universal. 

We obtain several other interesting results: (a) At short times the particles
move ballistically but, at large times, there is a crossover to Brownian
motion, at a crossover time $\Tcross$ that increases monotonically with $\St$.
(b) The PDFs $\mP(a)$, $\mP(\an)$, and $\mP(\at)$ all have exponential tails.
(c) By conditioning $\mP(\kappa)$ on the sign of the
Okubo-Weiss~\cite{oku70,wei91,per+ray+mit+pan11} parameter $\Lambda$, we show
that particles in regions of elongational flow ($\Lambda > 0$) have, on
average, trajectories with a lower curvature than particles in vortical regions
($\Lambda < 0$). 

We write the 2D incompressible Navier-Stokes (NS) equation in
terms of the stream-function $\psi$ and the vorticity $\bomega =
\curl \uu(\xx,t)$, where $\uu \equiv (-\partial_y \psi,
\partial_x\psi)$ is the fluid velocity at the point $\xx$ and
time $t$, as follows:
\begin{eqnarray}
D_t \bomega &=& \nu \nabla^2 \bomega
        - \mu \bomega  + F;
                \label{ns}\\
\nabla^2 {\bf \psi}  &=& \bomega.
            \label{poisson}
\end{eqnarray}
Here, $D_t\equiv\partial_t + \uu\cdot\nabla$, the uniform fluid
density $\rhof = 1$, $\mu$ is the coefficient of friction, and $\nu$
the kinematic viscosity of the fluid. We use a Kolmogorov-type
forcing $F(x,y)\equiv -F_0 \kinj \cos(\kinj y)$, with amplitude
$F_0$ and length scale $\linj\equiv 2\pi /\kinj$.  
(A) For $k<\kinj$, the inverse cascade of
energy yields $E(k) \sim k^{-5/3}$; and (B) for $k> \kinj$, there is a forward
cascade of enstrophy and $E(k) \sim k^{-\delta}$, where the exponent $\delta$
depends on the friction $\mu$ (for $\mu=0$, $\delta = 3$).
We use $\mu = 0.01$ and obtain $\delta = -3.6$.
The equation of motion for a small, spherical, rigid particle
(henceforth, a heavy particle) in an incompressible flow
~\cite{max+ril83} assumes the following simple form, if 
$\rhop \gg \rhof$ :
%
%
\begin{equation}
\ddt{\bf{x}} = \vv(t), \hspace{1cm}
\ddt{\vv} = - \frac{1}{\taus} \left[ \vv(t) -\uu(\bf{x}(t),t)  \right],
            \label{eq:heavy}
\end{equation}
where $\bf{x}$, $\vv$, and  $\taus = (2 \ap^2)\rhop/(9\nu\rhof)$ are,
respectively, the position, velocity, and response time of the particle, and 
$\ap$ is its radius.  We assume that $\ap \ll\eta$, the
dissipation scale of the carrier fluid, and that the particle number density is
so low that we can neglect interactions between particles, the particles
do not affect the flow, and particle accelerations are so high that we can
neglect gravity. In our DNSs we solve simultaneously for several species of
particles, each with a different value of $\St$; there are $\Np$ particles of
each species. We also obtain the trajectories for $\Np$ Lagrangian particles,
each of which obeys the  equation $ d(\bf{x})/dt =
\uu\left[\bf{x}(t),t\right]$. The details of our DNS are given in the
Appendix~\ref{suppli} and parameters in our DNSs are given in
Tables(\ref{table:para0}) and (\ref{table:para}) for $12$ representative values
of $\St$ (we have studied $20$ different values of $\St$). 

\begin{figure}
\includegraphics[width=0.95\columnwidth]{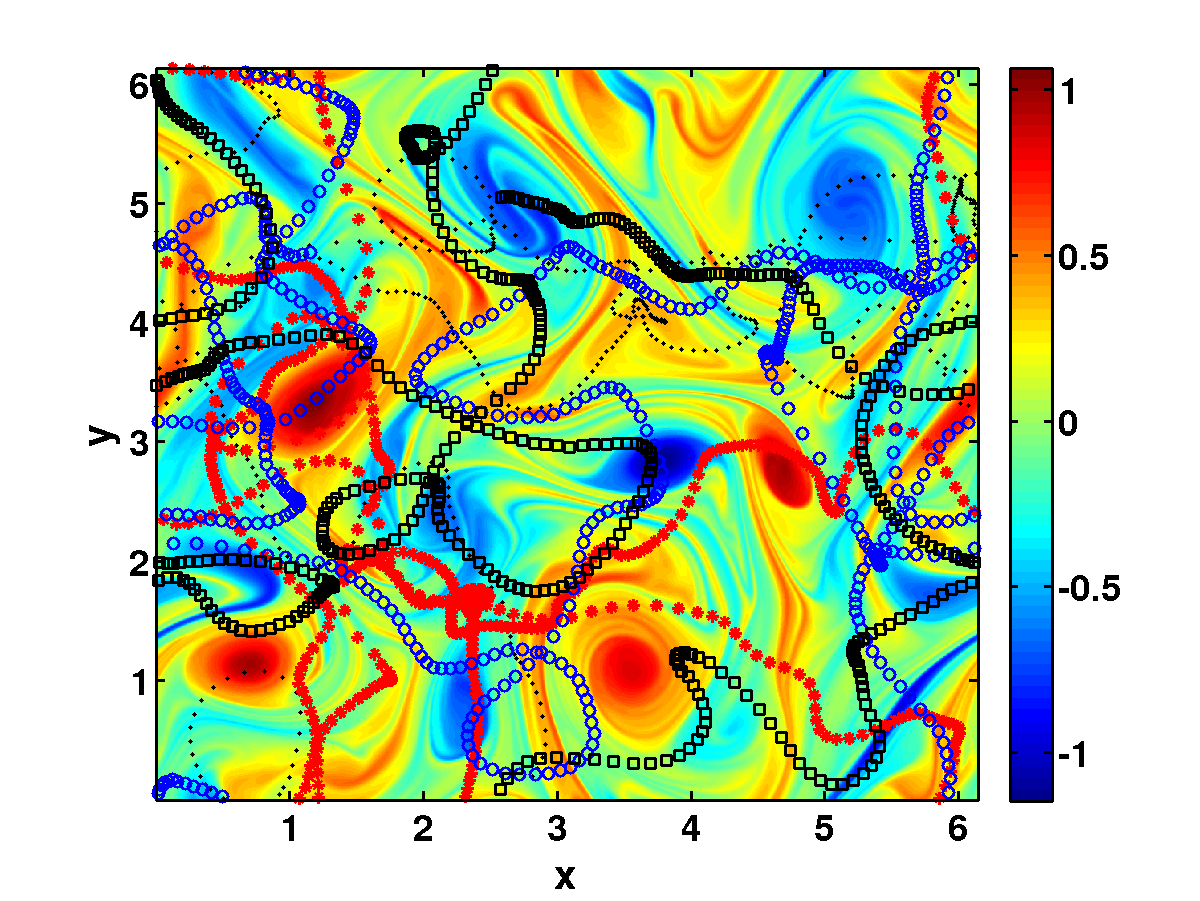}
\caption{\label{fig:track} (Color online)
Representative particle trajectories of a Lagrangian tracer (black line) and
three different heavy particles with $\St=0.1$ (red asterisks), $\St=0.5$ (blue
circles), and $\St=1$ (black squares) superimposed on a pseudocolor plot of
$\bomega$. For the spatiotemporal evolution of this plot see the 
animation available at the location \url{http://www.youtube.com/watch?v=lk3iSHhfTuU}
}
\end{figure}

\begin{figure}
\includegraphics[width=0.95\linewidth]{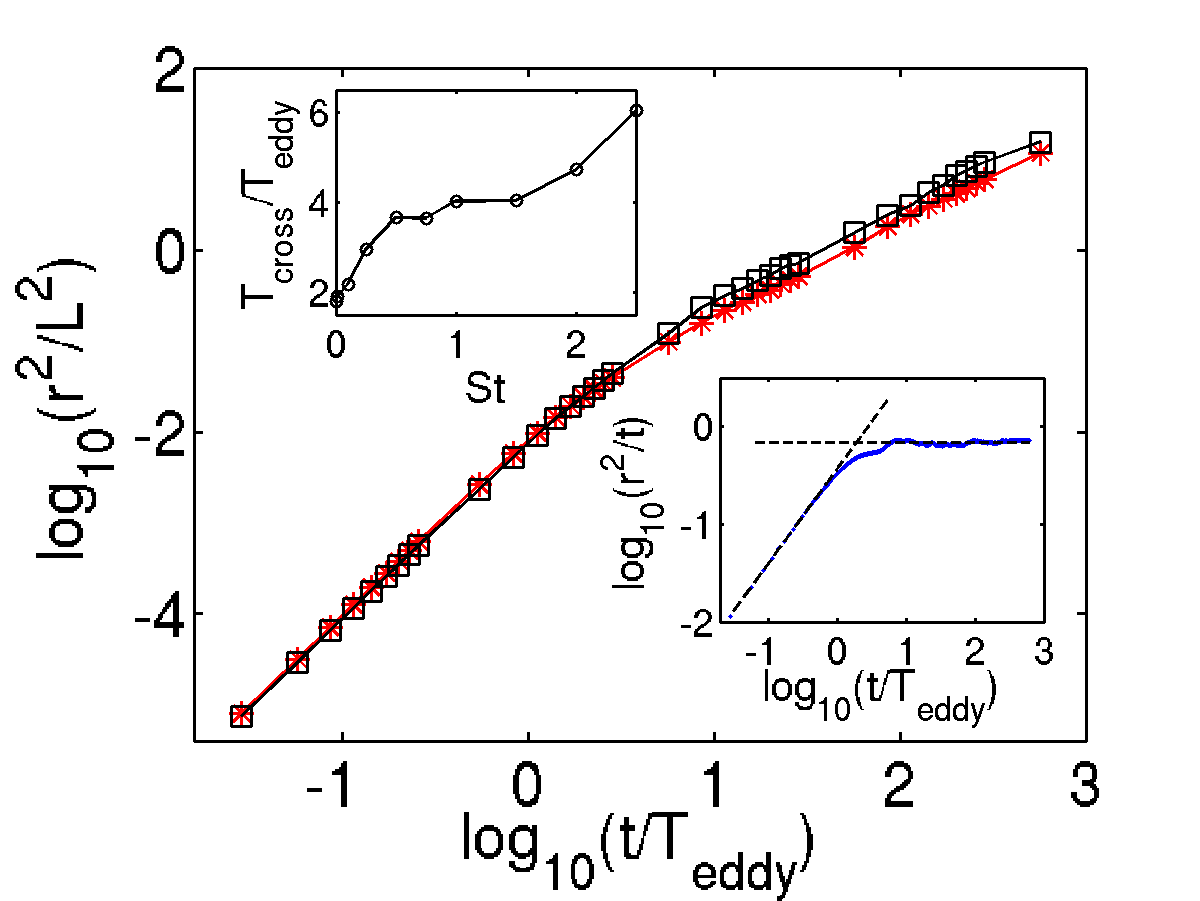}
\caption {\label{fig:msd} (Color online) Log-log (base 10) plots of $r^2$ versus $t/\Teddy$
for $\St=0.1$ (red triangles), and $\St=1$ (black squares); top inset: plot
of $\Tcross/\Teddy$ versus $\St$; bottom inset: log-log (base 10) plot of
$r^2/t$ versus $t/\Teddy$ for tracers (blue curve) and linear fits to the 
small- and large-$t$ asymptotes (dashed lines) with slopes $1$ and $0$
in ballistic and Brownian regimes, respectively; the intersection point
of these dashed lines yields $\Tcross$.}
\end{figure}

In \Fig{fig:track} we show  representative particle trajectories of a
Lagrangian tracer (black line) and three different heavy particles with
$\St=0.1$ (red asterisks), $\St=0.5$ (blue circles), and $\St=1$ (black
squares) superimposed on a pseudocolor plot of $\bomega$. 
We expect that inertial particles move ballistically in the range 
$0 < t \le \taus$; for $t \gg \taus$, we anticipate a crossover to 
Brownian behavior, which we quantify by defining the mean-square displacement 
$r^2(t) = \langle ({\bf x}(t_0+t)-{\bf x}(t_0))^2 \rangle_{t_0,\Np}$, 
where $\langle \rangle_{t_0,\Np}$ denotes an average over $t_0$ and over the
$\Np$ particles with a given value of $\St$. Figure (\ref{fig:msd}) contains
log-log plots of $r^2$ versus $t$, for the representative cases with $\St=0.1$
(red asterisks) and $\St=1$ (black squares); both of these plots show clear
crossovers from ballistic ($r^2 \sim t^2$) to Brownian ($r^2 \sim t$)
behaviors. We define the crossover time $\Tcross$ as the intersection of the
ballistic and Brownian asymptotes (bottom inset of \Fig{fig:msd}). The top
inset of \Fig{fig:msd} shows that, in the parameter range we consider, 
$\Tcross$ increases monotonically with $\St$.


\begin{figure*}
\includegraphics[width=0.32\linewidth]{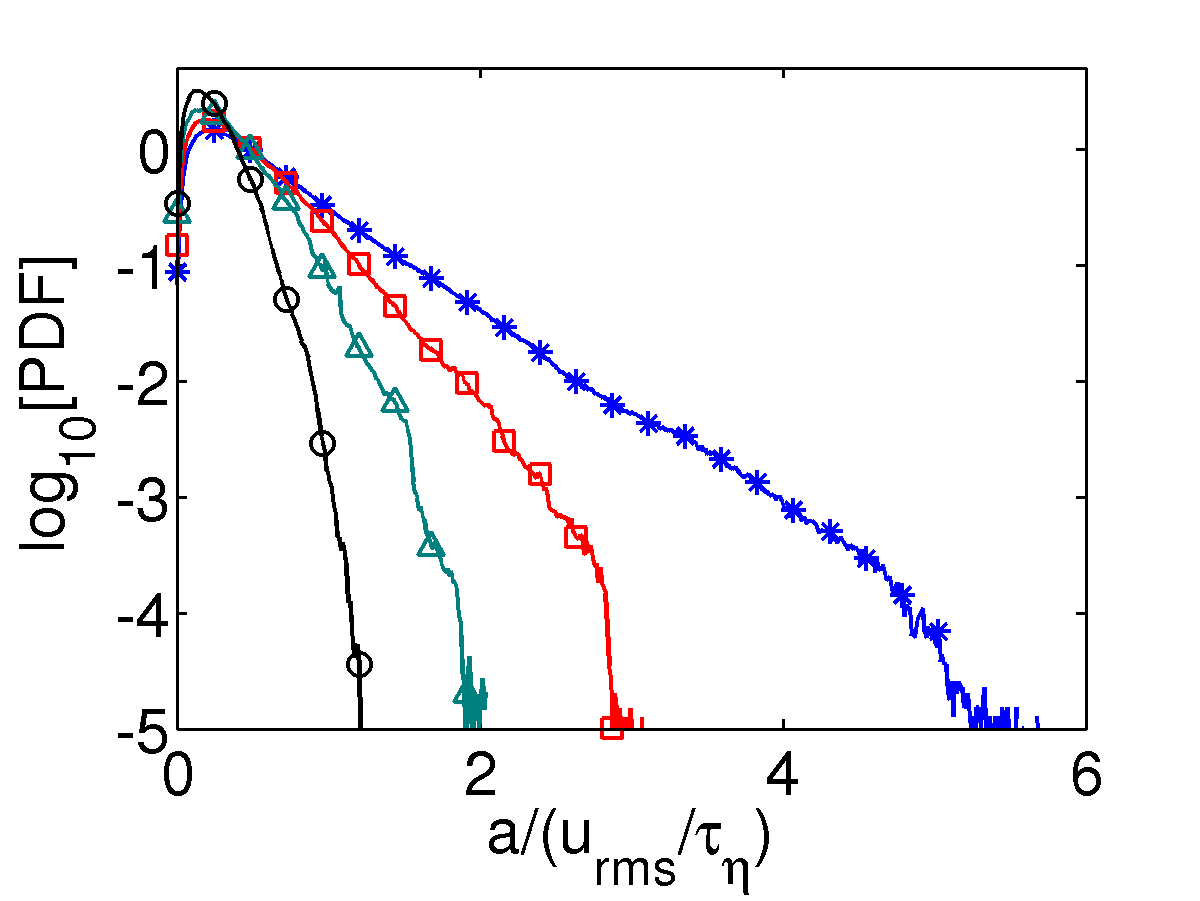}
\put(-45,85){ { {\bf (a)} } }
\includegraphics[width=0.32\linewidth]{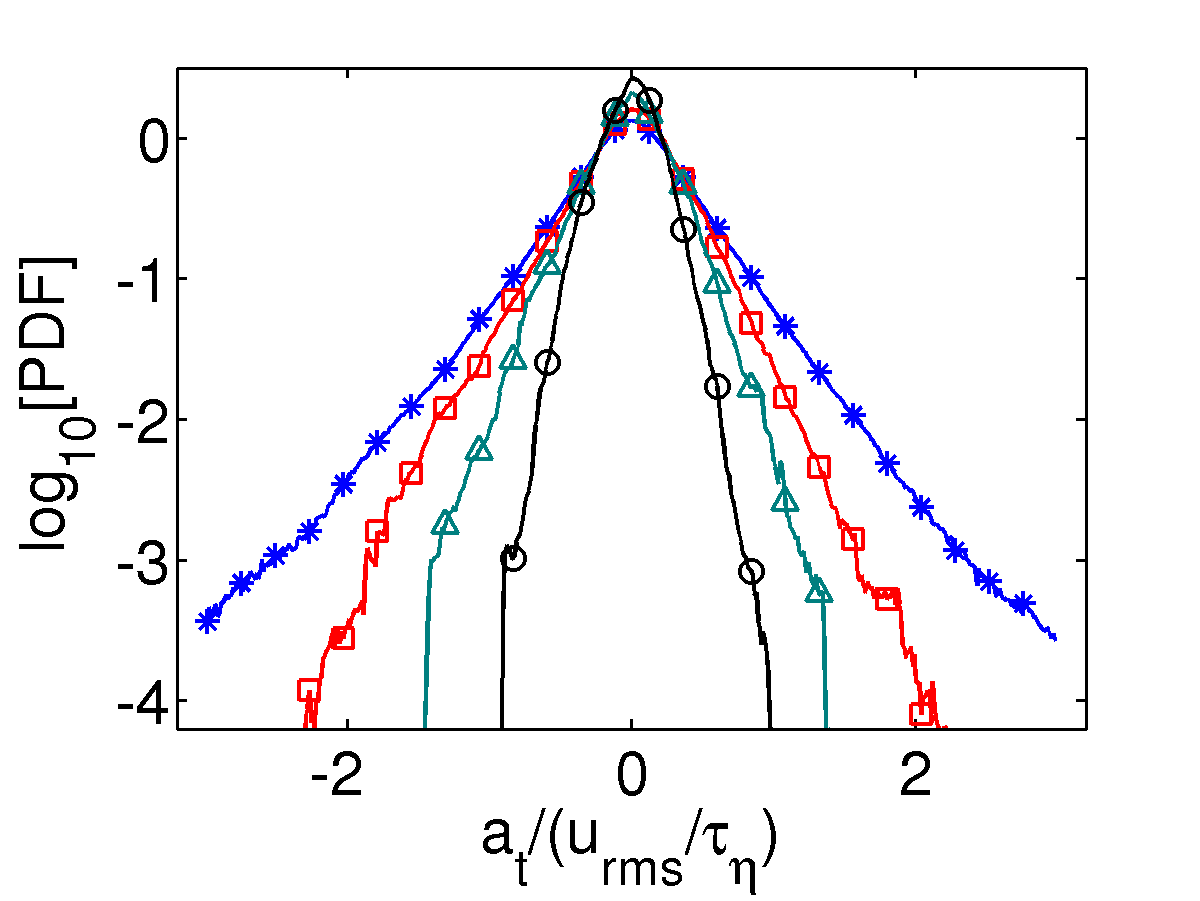}
\put(-45,85){ { {\bf (b)} } } 
\includegraphics[width=0.32\linewidth]{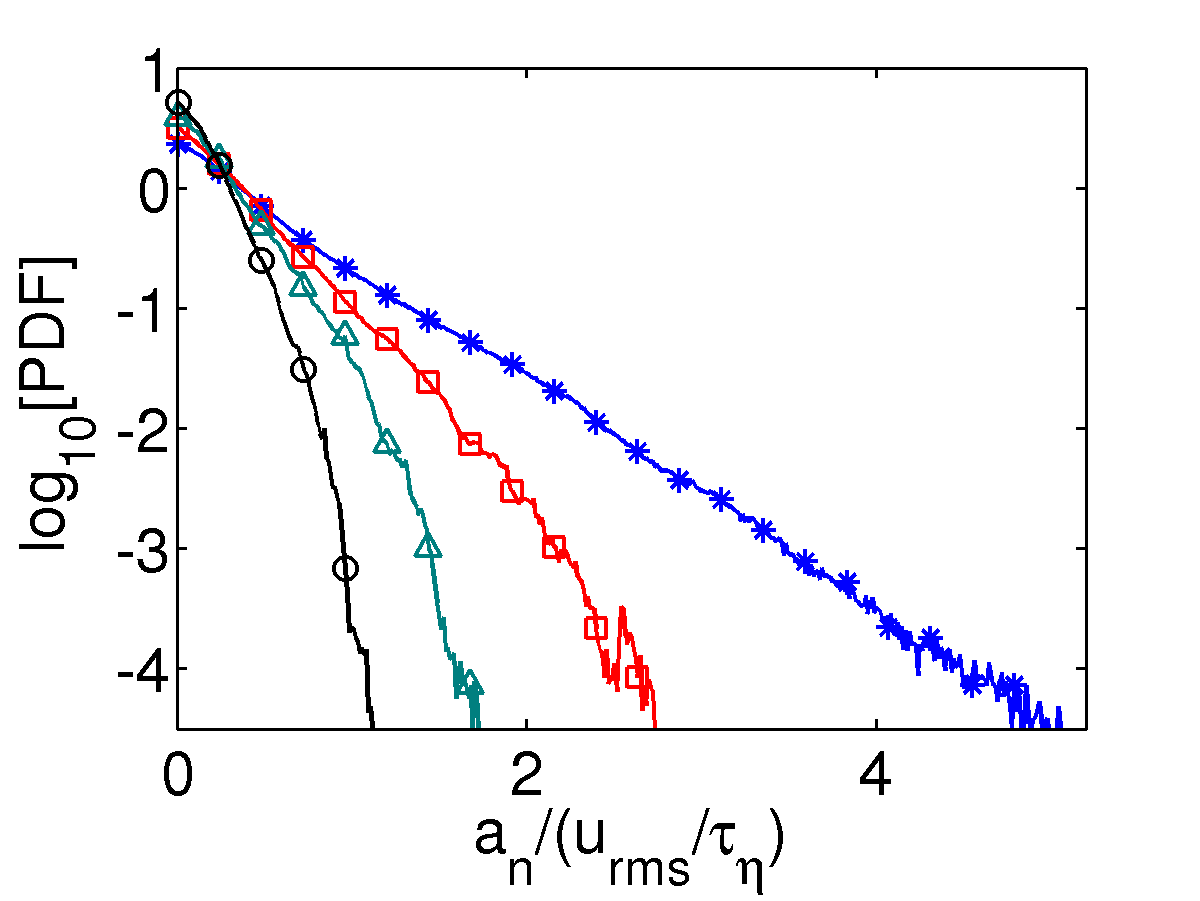}
\put(-45,85){ { {\bf (c)} } }
\caption {\label{fig:accn} (Color online) Plots of PDFs of (a) the modulus of $a$ of 
the particle acceleration, (b) its tangential component $\at$, and (c)
its normal component $\an$ for $\St= 0 $ (blue curve), $0.5$ (red curve), $1$ 
(green curve), and $2$ (black curve).
}
\end{figure*}

In \Fig{fig:accn} we present semilog plots of the PDFs $\mP(a)$, $\mP(\at)$,
and $\mP(\an)$ for some representative values of $\St$.  Clearly, all of
these PDFs have exponential tails, i.e., $\mP(a,\St) \sim \exp[-a/\alpha(\St)]
$, $\mP(\at,\St) \sim \exp[-\at/\alpha_{\rm t}(\St) ] $, and $\mP(\an,\St) \sim
\exp[-\an/\alpha_{\rm n}(\St) ] $.  As $\St$ increases, the tails of these PDFs
fall more and more rapidly, because the higher the inertia the more difficult
is it to accelerate a particle. Hence, $\alpha$, $\alphat$, and $\alphan$
decrease with $\St$ [see Table (\ref{table:para})].

\begin{figure}
\includegraphics[width=0.98\linewidth]{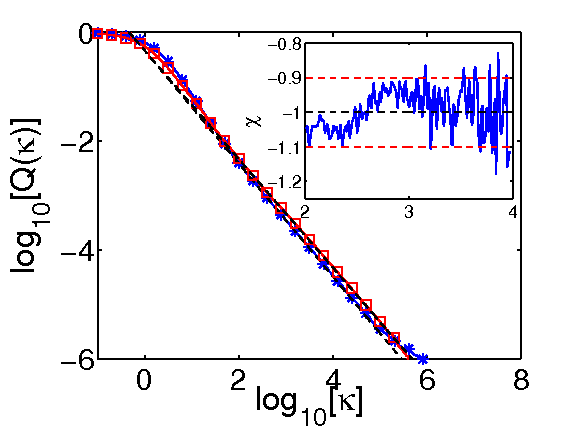}
\caption {\label{fig:curv} (Color online) Log-log plots of the cumulative PDFs 
$\mQ(\kappa)$ for $\St= 0.1$ (blue asterisks) and $\St=1$ (red squares); 
the inset shows a plot of the local slope of the tail of this cumulative 
PDF and the two dashed horizontal lines indicate the maximum and minimum 
values of this local slope in the range we use for fitting the exponent $\hr$. 
}
\end{figure}

Although these acceleration PDFs have exponential tails, $\mP(\kappa)$ shows a
power-law behavior as $\kappa\to \infty$, as we have mentioned above. The
exponent $\hr$ for the right-tail of $\mP(\kappa)$ is especially interesting
because it characterizes the parts of a trajectory that have large values of
$\kappa$. If $\mP(\kappa) \sim \kappa^{-\hr}$, then its cumulative PDF
$\mQ(\kappa) \sim \kappa^{-\hr +1}$.  We obtain an accurate estimate of $\hr$
from $\mQ$, which we obtain by a rank-order method that does not suffer from
binning errors~\cite{mit05a}.  We give representative, log-log plots of $\mQ$
in \Fig{fig:curv}, for $\St=0.1$ (blue asterisks) and $\St=1$ (red squares);
and we determine $\hr$ by fitting a straight line to $\mQ$ over a  scaling
range of more than two decades; We plot, in the inset, \Fig{fig:curv}, the
local slope of this scaling range, whose mean value and standard deviation
yield, respectively, $\hr$ and its error bars.  From such plots we find that
$\hr$ does not depend significantly on $\St$ [Table (\ref{table:para})].
Furthermore, we find that the Lagrangian analog of $\hr$, which we denote by
$h_{\rm lagrangian}$, is $2.03 \pm 0.09$, i.e., it lies within error bars of
$\hr$.  By analyzing the $\kappa \to 0$ limit of $\mP(\kappa)$, we find that
$\mP(\kappa) \sim A_0 \kappa^{\hll}$, where $A_0 > 0$ is an amplitude and $\hll
= 0.0 \pm 0.1$ (the latter is independent of $\St$); this 
indicates that there is a nonzero probability that the paths of particles 
have zero curvature, i.e., they can move in straight lines. The 
$\kappa \to 0$ limit of $\mP(\kappa)$ seems, therefore, to be different
from its counterpart for 3D fluid turbulence (see Ref.~\cite{xu+oue+bod07} for
Lagrangian tracers and Ref.~\cite{akshaypreprint} for heavy particles),
where $\mP(\kappa)\to 0$ as $\kappa \to 0$. Very-high-resolution DNSs for
2D turbulence must be undertaken to probe the $\kappa \to 0$ limit of 
$\mP(\kappa)$ by going to even smaller values of $\kappa$ than we
have been able to obtain reliably in our DNS. 

A point in a 2D flow is vortical or strain-dominated if the Okubo-Weiss
parameter $\Lambda = (1/8)(\omega^2 - \sigma^2)$ is, respectively, positive
or negative~\cite{oku70,wei91,per+ray+mit+pan11}. We now investigate how the
acceleration statistics of heavy particles depends on the sign of $\Lambda$ by
conditioning the PDFs of $\at$ and $\kappa$ on this sign. In particular, we
obtain the conditional PDFs $\mP^+$ and $\mP^-$, where the superscript stands
for the sign of $\Lambda$. We find, on the one hand, that the tail of
$\mP^{+}(\at)$ falls faster than that of $\mP^-(\at)$ because regions of the
trajectory with high tangential accelerations are associated with
strain-dominated points in the flow. On the other hand, the right tail of
$\mP^+(\kappa)$ falls more slowly than that of $\mP^-(\kappa)$, which implies
that high-curvature parts of a particle trajectory are correlated with vortical
regions of the flow. We give plots of $\mP^+(\at)$, $\mP^+(\kappa)$,
$\mP^-(\at)$, and $\mP^-(\kappa)$ in the Appendix~\ref{suppli}.
 
\begin{figure}
\includegraphics[width=0.95\columnwidth]{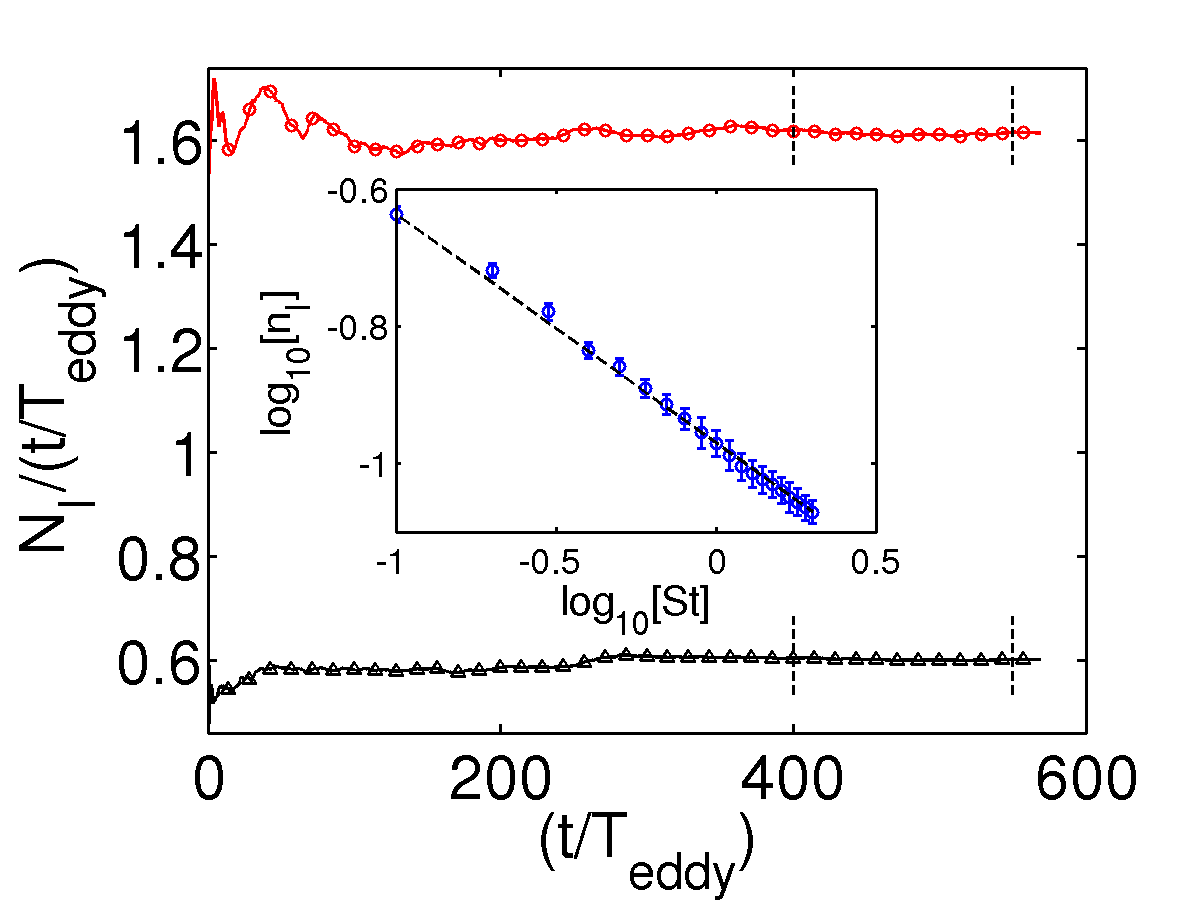}
\caption {\label{fig:inflection} (Color online) Plots of $\NI/(t/\Teddy)$ versus $t/\Teddy$ 
for $\St = 0.1$ (red curve) and $\St = 2$ (black curve); the inset shows a 
log-log (base 10) plot of $\nI$ versus $\St$ (blue open circles); the black 
dotted line has a slope $=-1/3$.}
\end{figure}

We find that $\aa \times \vv$ (a pseudoscalar in 2D like the
vorticity) changes sign at several \textit{inflection} points along a particle
trajectory. We use the number of inflection points on a trajectory, per unit
time, $\nI(\St)$ (see \Eq{eq:nI}) as a measure of its complexity.  In
\Fig{fig:inflection} we demonstrate that the limit in \Eq{eq:nI} exists by
plotting $\NI(t,\St)/t$ as a function of $t$ for $\St=0.1$ (red asterisks) and
$\St=2$ (black triangles); the mean value of $\NI(t,\St)/t$, between the two
vertical dashed lines in \Fig{fig:inflection}, yields our estimate for
$\nI(\St)$, which is given in the inset as a function of $\St$ (on a log-log
scale); the standard deviation gives the error bars.  From this inset of
\Fig{fig:inflection} we conclude that $\nI(\St) \sim \St^{-\Delta},$ with
$\Delta = 0.33 \pm 0.05$. This exponent $\Delta$ [Table~(\ref{table:para0})] is
independent of the Reynolds number and $\mu$, within the range of parameters we
have explored.  Furthermore, $\Delta$ is independent of whether our 2D
turbulent flow is dominated by forward or the inverse cascades in $E(k)$, which
are controlled by $\kinj$.

We have repeated all the above studies with a forcing term that yields an
energy spectrum with a significant inverse-cascade part ($\kinj = 50$); the
parameters for this run are given in Table (1) in the Appendix~\ref{suppli} and
in Ref.~\cite{AGthesis}.  The dependence of all the tails of the PDFs discussed
above and the exponents $\hll$ and $\hr$ on $\St$ are similar to those we have
found above for $\kinj = 4$.

Earlier studies of the geometrical properties of particle tracks
have been restricted to tracers; and they have inferred these properties
from tracer velocities and accelerations. For example, the
PDFs of different components of the acceleration of \textit{Lagrangian}
particles in 2D turbulent flows has been studied for both
decaying~\cite{wil+kam+fri08} and forced~\cite{kad+del+bos+sch11} cases; they
have shown exponential tails in periodic domains, but, in a confined
domain, have obtained PDFs with heavier tails~\cite{kad+bos+sch08}.  
The PDF of the curvature of tracer trajectories has been calculated 
from the same simulations, which quote an exponent $h_{\rm lagrangian} \simeq
2.25$ (but no error bars are given). Our work goes well beyond these
earlier studies by (a) investigating the statistical properties of the
geometries of the trajectories of {\it {heavy particles}} in 2D turbulent flows
for a variety of parameter ranges and Stokes numbers, (b) by introducing and
evaluating, with unprecedented accuracy (and error bars), the exponent $\hr$,
(c) proposing $\nI$ as a measure of the complexity of heavy-particle
trajectories and obtaining the exponent $\Delta$ accurately, (d) by
examining the dependence of all these exponents on $\St$ and $\kinj$, and (e)
showing, thereby, that these exponents are universal (within our error bars).

Our results imply that $\nI(\St)$ has a power-law divergence, so the
trajectories become more and more contorted, as $\St \to 0$.  This divergence
is suppressed eventually, in any DNS, which can only achieve a finite value of
$Re_{\lambda}$ because it uses only a finite number of collocation points. Such
a suppression is the analog of the finite-size rounding off of divergences, in
say the susceptibility, at an equilibrium critical point~\cite{fssprivman}.
Note also that the limit $\St \to 0$ is singular and it is not clear {\it a
priori} that this limit should yield the same results, for the properties we
study, as the Lagrangian case $\St =0$.

We hope that our study will lead to experimental studies and accurate
measurements of the exponents $\hr$ and $\Delta$, and applications of
these in developing a detailed understanding of particle-laden flows in the
variety of systems that we have mentioned in the introduction. 

For 3D turbulent flows, geometrical properties of Lagrangian-particle
trajectories have been studied numerically~\cite{bra+lil+eck06,sca11} and
experimentally~\cite{xu+oue+bod07}. However, such geometrical properties have
not been studied for heavy particles. The extension of our heavy-particle study
to the case of 3D fluid turbulence is nontrivial and will be given in a
companion paper~\cite{akshaypreprint}.

\begin{table}
   \begin{tabular}{@{\extracolsep{\fill}} c c c c c c c c c c c c c}
    \hline
    $Run $ &$N$ & $F_0$ & $\kinj$ & $\ld$ & $\lambda$ & $Re_{\lambda}$ &
$\Teddy$ & $\Teta$ & $\Tinj$\\
   \hline \hline
    {\tt IA}  & $1024$ & $0.2$ & $50$ &  $1.3 \times 10^{-3}$	&
$0.06$ & $1219$ & $0.98$ & $0.16$ & $2.94$\\
{\tt FA}  & $1024$ & $0.005$ & $4$ &  $5.4 \times 10^{-3}$ &
$0.2$ & $1322$ & $7$ & $2.9$ & $30.2$\\

\hline
\end{tabular}

\caption{\small
The parameters for our DNS runs: $N^2$ is the number of collocation points,
$\Np = 10^4$ is the number of Lagrangian or inertial particles, $\delta t$ the
time step, $\nu = 10^{-5}$ the kinematic viscosity, and $\mu = 0.01$ the
air-drag-induced friction, $F_0$ the forcing amplitude, $\kinj$ the forcing
wave number, $l_d \equiv (\nu^3/\varepsilon)^{1/4}$ the dissipation scale,
$\lambda \equiv \sqrt {\nu E/\varepsilon}$ the Taylor microscale, $Re_{\lambda}
= \urms \lambda/\nu$ the Taylor-microscale Reynolds number, $T_{eddy} =
(\frac{\sum_k E(k)/k}{\sum_k E(k)})/u_{rms}$ the eddy-turn-over time, and
$\Teta \equiv \sqrt{(\nu/\varepsilon)}$ the Kolmogorov time scale.
$\Tinj\equiv (\linj^2/E_{\rm inj})^{1/3}$ is the energy-injection time scale,
where $E_{\rm inj}=<{\bf f_{\rm u}}\cdot \uu>$, is the energy-injection rate,
$\linj=2\pi/\kinj$ is the energy-injection length scale, and ${\bm f}_{\omega}
= \nabla \times {\bm f}_{\rm u}$.  } \label{table:para0} \end{table} 
\begin{table}
 \begin{tabular}{@{\extracolsep{\fill}} c c c c c c c c }
    \hline
    $Run $ &  $\St$ & $\alpha$     &    $\alphat$     &   $\alphan$    & $\hr$ \\
   \hline \hline
   F1        & $0.1$  & $0.86 \pm0.07$ &    $1.45\pm0.07$   &   $0.86 \pm0.07$  
& $ 2.03\pm 0.08$ \\   
   F2        & $0.2$  & $0.96 \pm0.06$ &    $1.66\pm0.07$   &   $0.97 \pm0.06$  
& $ 2.0\pm 0.1$ \\   
   F3        & $0.3$  & $1.11 \pm0.07$ &    $1.87\pm0.07$   &   $1.12 \pm0.06$  
& $ 2.0\pm 0.1$ \\   
   F4        & $0.4$  & $1.43 \pm0.07$ &    $2.15\pm0.07$   &   $1.36 \pm0.09$  
& $ 2.04\pm 0.09$ \\   
   F5        & $0.5$  & $1.56 \pm0.08$ &    $2.27\pm0.08$   &   $1.45 \pm0.09$  
& $ 2.0\pm 0.1$ \\ 
   F6        & $0.6$  & $1.66 \pm0.08$ &    $2.36\pm0.09$   &   $1.6 \pm0.1$  
& $ 2.02\pm 0.09$ \\   
   F7        & $0.7$  & $1.88 \pm0.09$ &    $2.51\pm0.09$   &   $1.61 \pm0.09$  
& $ 2.06\pm 0.09$ \\   
   F8        & $0.8$  & $2.22 \pm0.08$ &    $2.73\pm0.09$   &   $1.90 \pm0.09$  
& $ 2.01\pm 0.08$ \\   
   F9        & $0.9$  & $2.6 \pm0.1$ &    $2.9\pm0.1$   &   $2.0 \pm0.1$  
& $ 2.0\pm 0.1$ \\   
   F10      & $1.0$  & $2.6 \pm0.1$ &    $3.3\pm0.1$   &   $2.17 \pm0.09$  
& $ 2.0\pm 0.1$ \\   
   F11      & $1.5$  & $3.9 \pm0.1$ &    $4.3\pm0.1$   &   $3.3 \pm0.1$  
& $ 2.1\pm 0.1$ \\   
   F12      & $2.0$  & $4.5 \pm0.1$ &    $4.7\pm0.1$   &   $3.8 \pm0.1$  
& $ 2.0\pm 0.1$ \\   
\hline
\end{tabular}
\caption{\small
The values of $\alpha$, $\alphan$, and $\alphat$ and the exponent $\hr$ for the case $\kinj =4$ and for different values of $\St$. 
}
\label{table:para}
\end{table} 

\section{Acknowledgments}

We thank A. Bhatnagar, A. Brandenburg, B. Mehlig, S.S. Ray, and D. Vincenzi for
discussions, and particularly A. Niemi, whose study of the intrinsic
geometrical properties of polymers~\cite{poly11}, inspired our work on particle
trajectories.  The work has been supported  in part by the European Research
Council under the AstroDyn Research Project No.\ 227952 (DM), Swedish Research
Council under grant 2011-542 (DM), NORDITA visiting PhD students program (AG),
and CSIR, UGC, and DST (India) (AG and RP).  We thank SERC (IISc) for providing
computational resources. AG, PP, and RP thank NORDITA for hospitality; DM
thanks the Indian Institute of Science for hospitality.


\appendix*
\section{Statistical Properties of the Intrinsic Geometry of 
Heavy-particle Trajectories in Two-dimensional, Homogeneous,
Isotropic Turbulence : Supplemental Material}
\label{suppli}

\begin{figure}[h]
\centering
\includegraphics[width=0.93\columnwidth]{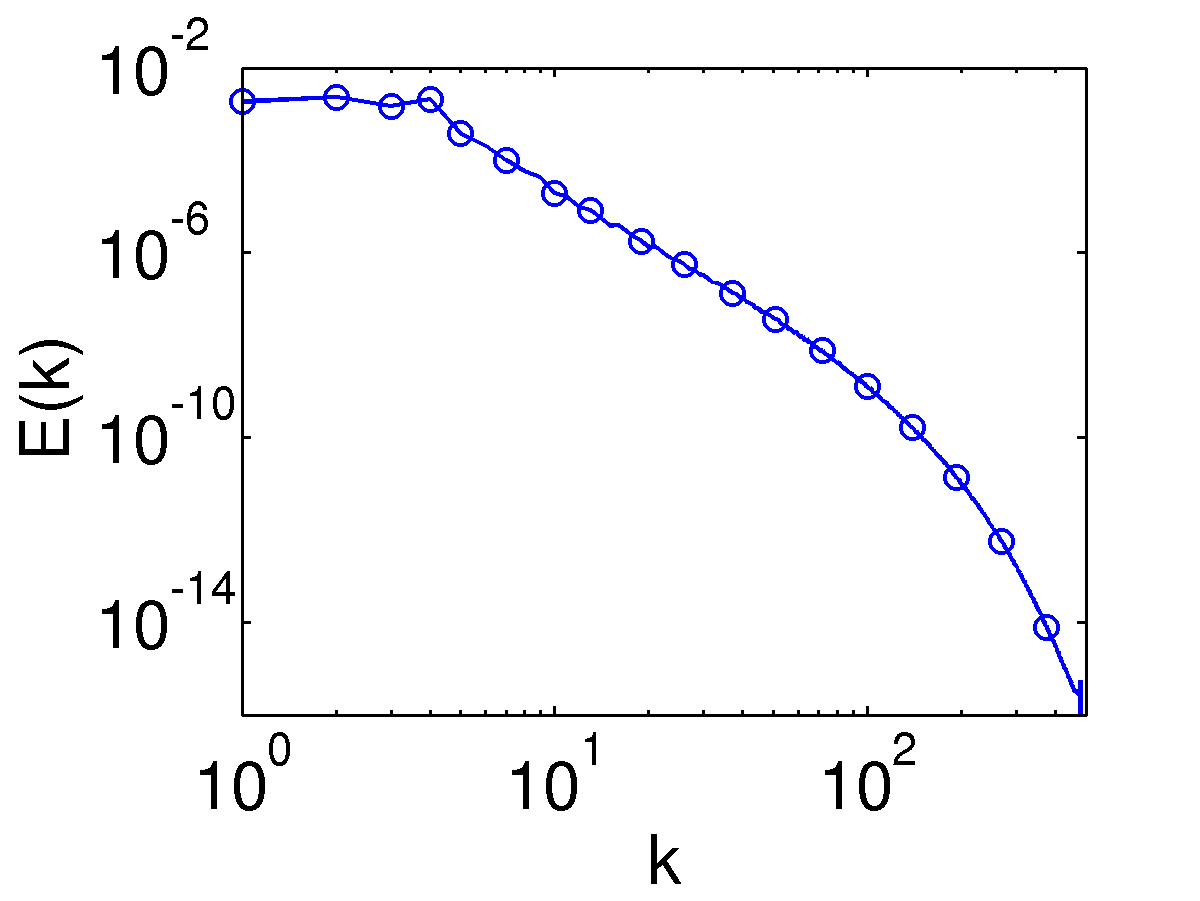}
\put(-55,135){ { {\Large (a)} } }\\
\includegraphics[width=0.93\columnwidth]{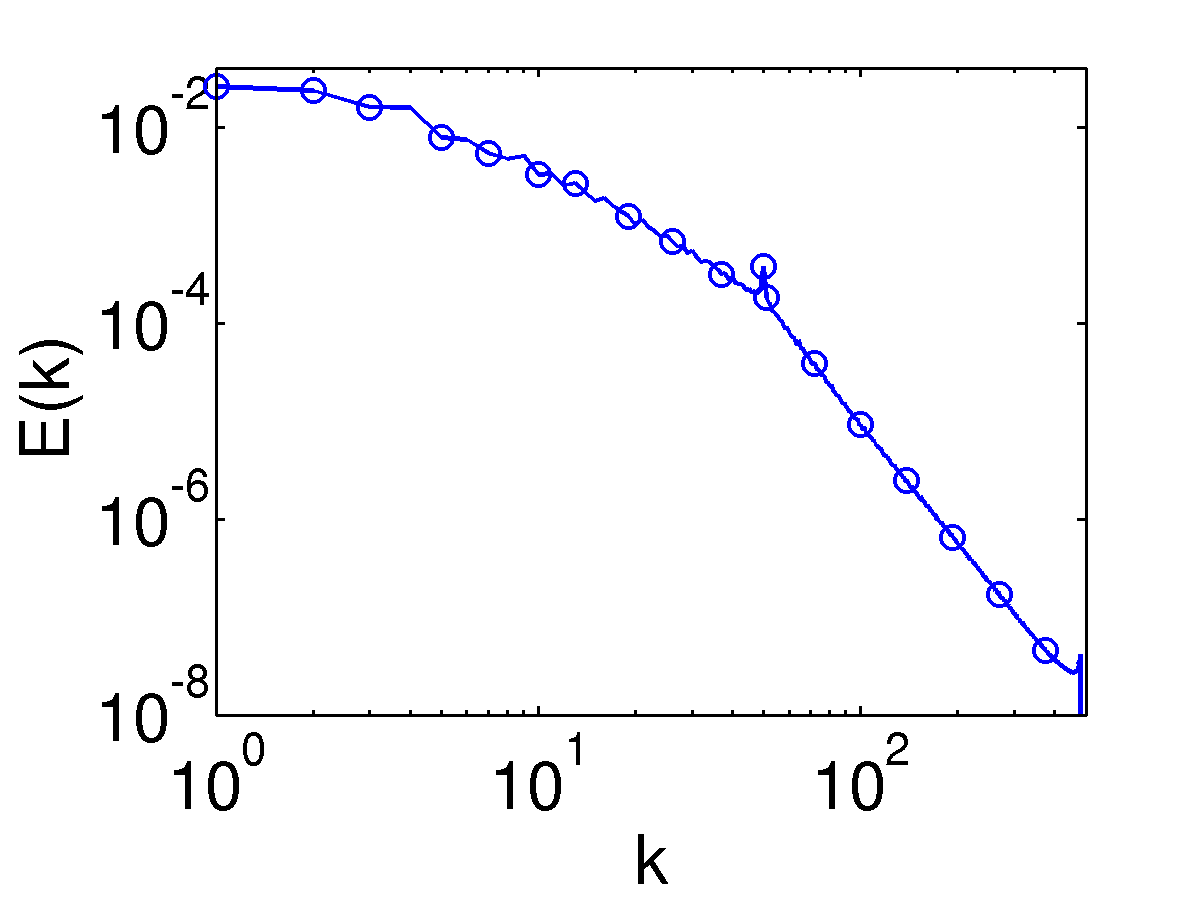}
\put(-55,135){ { {\Large (b)} } }
\caption{(Color online)
Log-log (base 10) plots of the energy spectra $E(k)$ versus $k$
for (a) runs $\tt FA$ ($\kinj = 4$) and (b) runs $\tt IA$ ($\kinj = 50$).
}
\label{fig:spectra}
\end{figure}

In this Supplemental Material we provide numerical details of our direct
numerical simulation (DNS) of Eq. (2) in the main part of this paper. We also
give results of our DNS for the case of the injection wave vector $\kinj = 50$,
which yields a significant inverse-cascade part in the energy spectrum $E(k)$.
In \Fig{fig:spectra} we show the energy spectra $E(k)$ for our runs $\tt FA$ 
 ($\kinj = 4$) and $\tt IA$ ($\kinj = 50$).

We perform a DNS of Eq. (2) by using a pseudo-spectral code~\cite{Can88} with
the $2/3$ rule for dealiasing; and we use a second-order, exponential time
differencing Runge-Kutta  method~\cite{cox+mat02} for time stepping.  We use
periodic boundary conditions in a square simulation domain with side $\mathbb L
= 2 \pi$, with $N^2$ collocation points.  Together with Eq.(2) we solve for the
trajectories of $\Np$ heavy particles, for each of which we solve Eq. (4) with
an Euler scheme. The use of an Euler scheme to evolve particles is justified
because, in time $\delta t$, a particle crosses at most one-tenth of grid
spacing.  We obtain the Lagrangian velocity at an off-grid particle position
$\xx$, from the Eulerian velocity field by using a bilinear-interpolation
scheme~\cite{Pre+Fla+Teu+Vet92}; for numerical details see
Refs.~\cite{per+pan09,Per09,per+ray+mit+pan11,ray+mit+per+pan11}.  

We calculate the fluid energy-spectrum $E(k)\equiv\sum_{k-1/2< k' \le k+1/2}
k'^2 \langle |\hat {\psi}({\bf k'},t)|^2 \rangle _t$, where $\langle \cdot
\rangle_t$ indicates a time average over the statistically steady state.  The
parameters in our simulations are given in Table(II) of the main part of this
paper and in Table(\ref{table:inverse}). These include the Taylor-microscale
Reynolds number, $Re_{\lambda} \equiv \urms \lambda/\nu$, where  $\lambda
\equiv \sqrt {\nu E/\varepsilon}$ is the Taylor microscale and the Stokes
number $\St = \taus/\Teta$.  We use $20$ different values of $\St$ to study the
dependence on $\St$ of the PDFs $\mP(a)$, $\mP(\at)$ and $\mP(\an)$, the
cumulative PDF $\mQ(\kappa)$, the mean square displacement, and the number of
inflection points $\NI(t,\St)$ at which $\aa \times \vv$ changes sign along a
particle trajectory.

\begin{figure}[h]
\centering
\includegraphics[width=0.93\columnwidth]{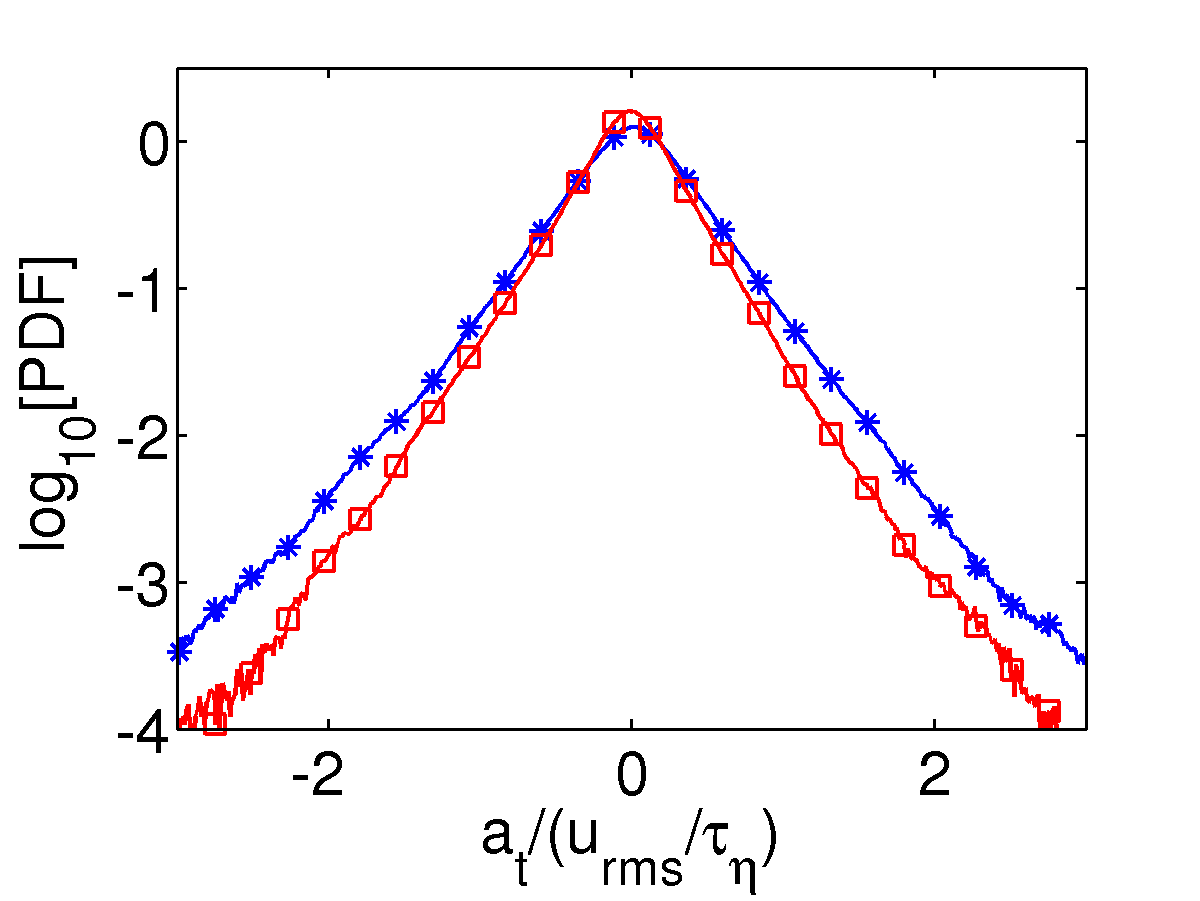}
\caption{(Color online) 
Semilog (base 10) plots of the PDFs of the tangential component of 
the acceleration for $\St=0.1$ in vortical regions $\mP(\at^+)$ (red squares) 
and in strain-dominated regions $\mP(\at^-)$ (blue asterisks).
}
\label{fig:pdf-acceleration}
\end{figure}

\begin{figure}[h]
\centering
\includegraphics[width=0.95\columnwidth]{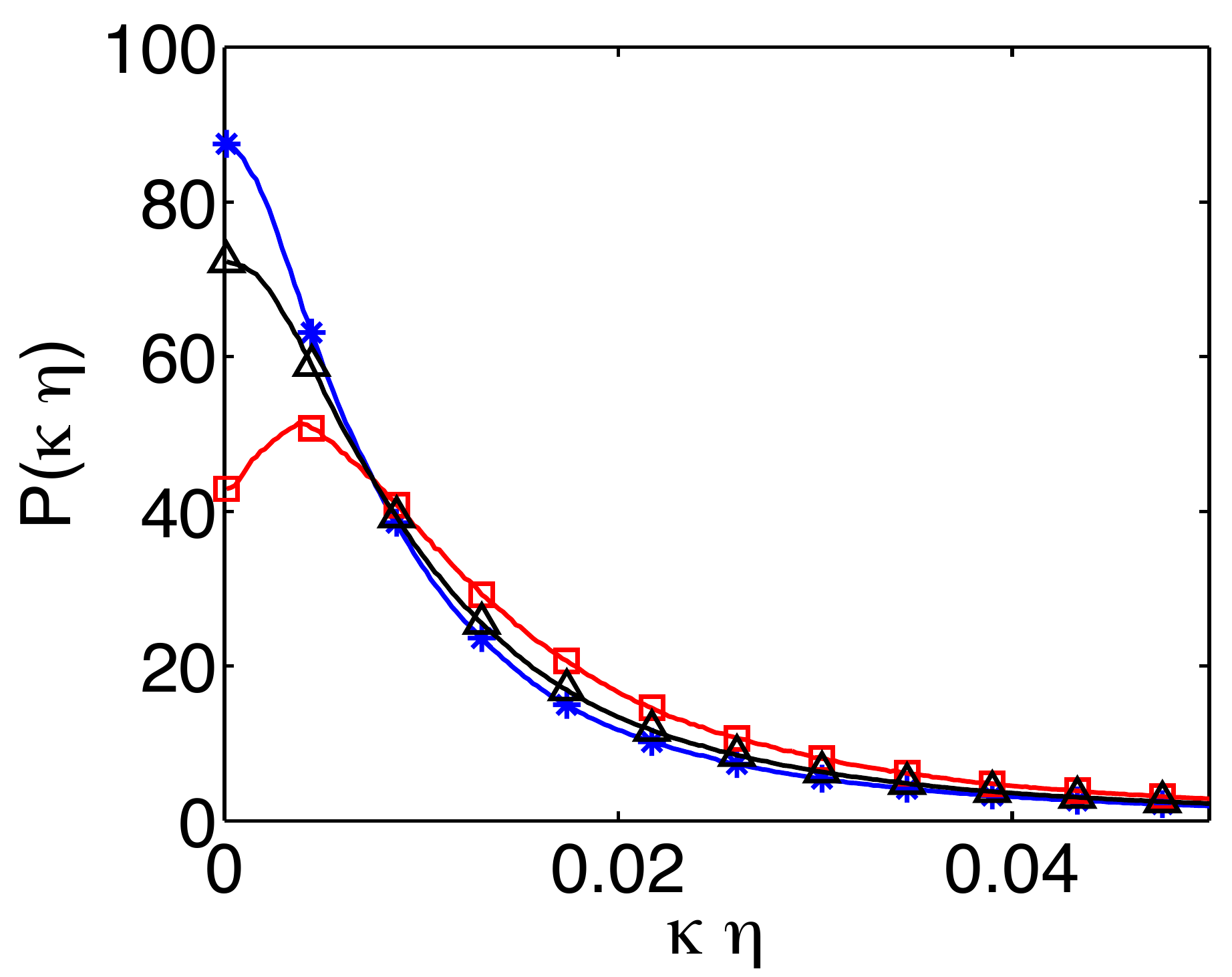}%
\caption{(Color online) 
Semilog (base 10) plots of PDF of the curvature of trajectories for $\St=0.1$ 
in vortical regions $\mP(\kappa^+ \eta)$ (red squares), in strain-dominated 
regions $\mP(\kappa^- \eta)$ (blue asterisks), and in general (i.e., without 
conditioning on the sign of $\Lambda$) $\mP(\kappa \eta)$ (black triangles).  
}
\label{fig:pdf-kappa}
\end{figure}

A point in a 2D flow is vortical or strain-dominated if the Okubo-Weiss
parameter $\Lambda = (1/8)(\omega^2 - \sigma^2  )$ is, respectively, positive
or negative~\cite{oku70,wei91,per+ray+mit+pan11}.  We investigate how the
acceleration statistics of heavy particles depends on the sign of $\Lambda$ by
conditioning the PDFs of $\at$ and $\kappa$ on this sign. In particular, we
obtain the conditional PDFs $\mP^+$ and $\mP^-$, where the superscript stands
for the sign of $\Lambda$. We find, on the one hand, that the tail of
$\mP^{+}(\at)$ falls faster than that of $\mP^-(\at)$ because regions of the
trajectory with high tangential accelerations are associated with
strain-dominated points in the flow. On the other hand, the right tail of
$\mP^+(\kappa)$ falls more slowly than that of $\mP^-(\kappa)$, which implies
that high-curvature parts of a particle trajectory are correlated with vortical
regions of the flow. We give plots of $\mP^+(\at)$, $\mP^+(\kappa)$,
$\mP^-(\at)$, and $\mP^-(\kappa)$ in \Fig{fig:pdf-acceleration} and
\Fig{fig:pdf-kappa}. These trends hold for all values of $\St$ and $\kinj$ that
we have studied. 

\hspace*{1cm}
\hspace*{1cm}
\begin{figure}[h]
\centering
\includegraphics[width=0.95\columnwidth]{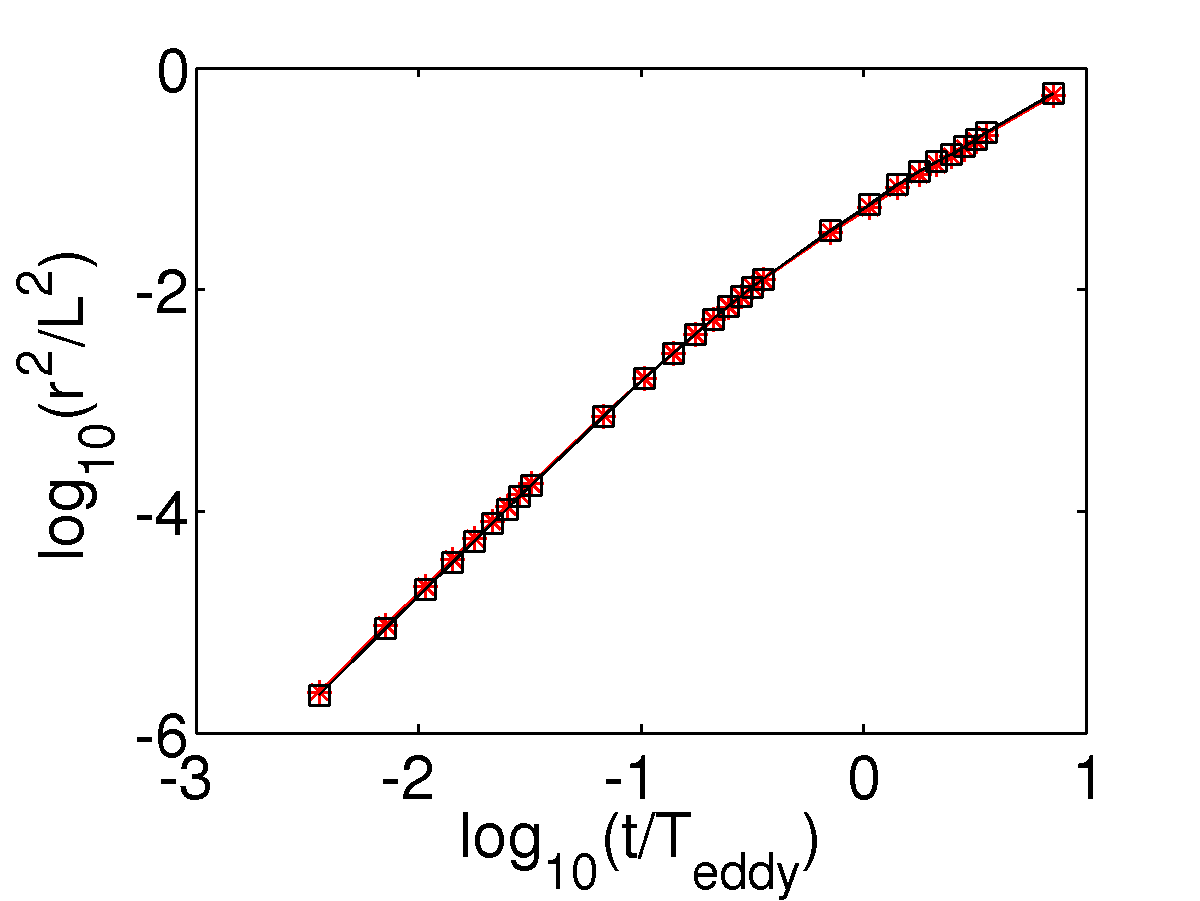}%
\caption{(Color online) 
Log-log (base 10) plots for $\kinj=50$ of $r^2$ versus $t/\Teddy$ for 
$\St=0.1$ (red asterisks) and $\St=1$ (black squares).
}
\label{fig:inv-msd}
\end{figure}

In \Fig{fig:inv-msd}, we plot the square of the  mean-squared
displacement $r^2$ versus time $t$ for $\kinj = 50$; here too we see a
crossove from ballistic to Brownian behaviors; however, in contrast to the
case $\kinj = 4$, the crossover time $\Tcross$ does not depend 
significantly on $\St$.
\hspace*{1cm}
\begin{figure}[h]
\centering
\includegraphics[width=0.95\linewidth]{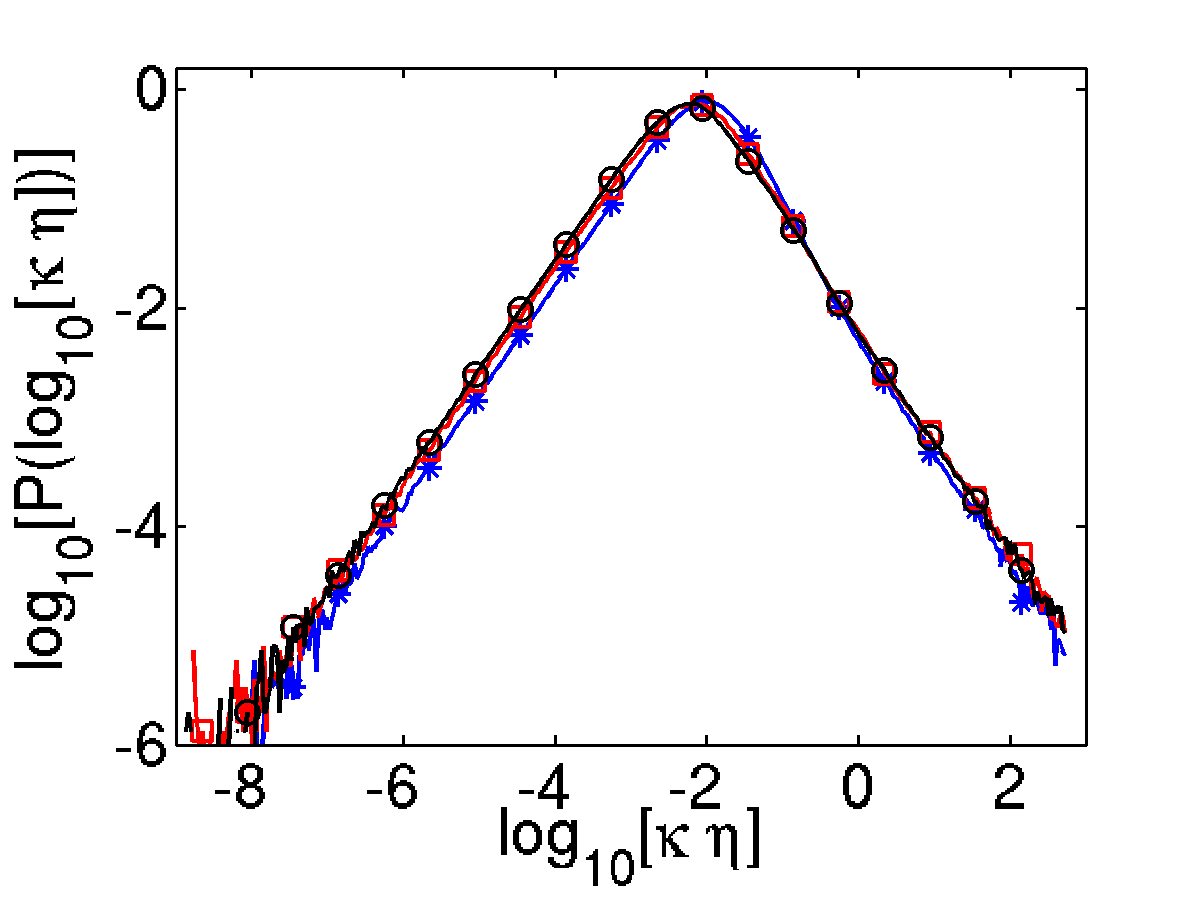}
\caption{(Color online) Semilog (base 10) plot of the PDF $\mP(\log_{10}(\kappa\eta))$ versus $\log_{10}(\kappa \eta)$, for $\St= 0.1$ (blue asterisks),   
301 $\St=1$ (red squares) and $\St=2$ (black circles).
}
\label{fig:log-curv}
\end{figure}

In \Fig{fig:log-curv}, we plot the PDF $\mP(\log_{10}(\kappa\eta))$ versus
$\log_{10}(\kappa \eta)$, for $\St= 0.1$ (blue asterisks), $\St=1$ (red
squares) and $\St=2$ (black circles). Such PDFs provide another convenient way
of displaying the power-law behaviors, as $\kappa\to \infty$ and $\kappa \to
0$, which we have reported in the main part of this paper, where we have used
the cumulative PDF of $\kappa$ to obtain the power-law exponents.


In Table(\ref{table:inverse}) we report the values of $\alpha$, $\alphan$,
$\alphat$, and the exponent $\hr$ of the right tail of the PDF of the
trajectory curvature, for the case $\kinj=50$ and for different values of
$\St$.

\begin{table}
 \begin{tabular}{@{\extracolsep{\fill}} c c c c c c c c }
    \hline
    $Run $ &  $\St$ & $\alpha$     &    $\alphat$     &   $\alphan$    & $\hr $ \\
   \hline \hline
   I1  & $0.1$  & $0.39 \pm 0.06$ &    $0.69\pm 0.02$   &   $0.40 \pm 0.06$  & $ 2.16\pm 0.09$ \\   
   I2  & $0.2$  & $0.47 \pm 0.05$ &    $0.81\pm 0.03$   &   $0.46 \pm 0.05$  & $ 2.14\pm 0.09$ \\   
   I3  & $0.3$  & $0.55 \pm 0.04$ &    $0.95\pm 0.02$   &   $0.54 \pm 0.05$  & $ 2.1\pm 0.1$ \\   
   I4  & $0.4$  & $0.63 \pm 0.04$ &    $1.09\pm 0.03$   &   $0.61 \pm 0.04$  & $ 2.10\pm 0.08$ \\   
   I5  & $0.5$  & $0.71 \pm 0.04$ &    $1.21\pm 0.02$   &   $0.68 \pm 0.03$  & $ 2.09\pm 0.09$ \\
   I6  & $0.6$  & $0.80 \pm 0.03$ &    $1.34\pm 0.03$   &   $0.77 \pm 0.03$  & $ 2.08\pm 0.09$ \\   
   I7  & $0.7$  & $0.88 \pm 0.04$ &    $1.48\pm 0.04$   &   $0.85 \pm 0.03$  & $ 2.07\pm 0.09$ \\   
   I8  & $0.8$  & $0.97 \pm 0.03$ &    $1.60\pm 0.03$   &   $0.94 \pm 0.04$  & $ 2.07\pm 0.09$ \\   
   I9  & $0.9$  & $1.05 \pm 0.03$ &    $1.73\pm 0.03$   &   $1.01 \pm 0.04$  & $ 2.1\pm 0.1$ \\   
   I10 & $1.0$  & $1.16 \pm 0.03$ &    $1.87\pm 0.03$   &   $1.10 \pm 0.03$  & $ 2.1\pm 0.1$ \\   
\hline
\end{tabular}
\caption{\small
The values of $\alpha$, $\alphan$, $\alphat$, and the exponent $\hr$, for 
the case $\kinj = 50$ for different values of $\St$. 
}
\label{table:inverse}
\end{table} 

In Table(\ref{table:parah}) we report the exponent $\hll$, which charcterizes
$\mP(\kappa \eta)$, as $\kappa \to 0$, in both the cases $\kinj = 4$ and $\kinj
= 50$.  In both these cases and for all the different values of $\St$ we have
studied, $\hll = 0.0 \pm 0.1$.


\begin{table}
   \begin{tabular}{@{\extracolsep{\fill}} c c c c c c c c c c c c c}
    \hline
    $\St$ &$0.1$ & $0.2$ & $0.3$ & $0.4$ & $0.5$ & $1.0$  \\
   \hline \hline
    {$\hll$ ($\tt FA$)}  & $0.0\pm0.1$ & $0.0\pm 0.1$ & $0.0 \pm 0.1$ &  $0.0 \pm 0.1$ & $0.0 \pm 0.1$
& $0.0 \pm 0.1$ \\
    {$\hll$ ($\tt IA$)}  & $0.0\pm0.1$ & $0.0\pm 0.1$ & $0.0 \pm 0.1$ &  $0.0 \pm 0.1$ & $0.0 \pm 0.1$
& $0.0 \pm 0.1$ \\
\hline
\end{tabular}

\caption{\small
The exponent $\hll$ that charcterizes $\mP(\kappa \eta)$, as $\kappa \to 0$, in
both the cases $\kinj = 4$ and $\kinj = 50$ and for different values of $\St$. 
} 
\label{table:parah}
\end{table} 
\end{document}